\documentclass[prl,twocolumn,superscriptaddress,10pt,showpacs]{revtex4-1}
\usepackage{graphicx}
\usepackage{color}
\begin{document}
\title{Interfacial free energy of a hard-sphere fluid  in contact with curved hard  surfaces}

\author{Brian B. Laird}
\thanks{Author to whom correspondence should be addressed}
\email{blaird@ku.edu}
\author{Allie Hunter}
\affiliation{Department of Chemistry,
University of Kansas,
Lawrence, KS 66045, USA}
\author{Ruslan L. Davidchack}
\affiliation{Department of Mathematics, University of Leicester, 
Leicester, LE1 7RH, UK}

\date{\today}
\begin{abstract}
Using molecular-dynamics simulation, we have calculated the interfacial free energy, $\gamma$, between a hard-sphere fluid and hard spherical and cylindrical colloidal particles, as functions of the  particle radius $R$ and the fluid packing fraction $\eta = \rho\sigma^3/6$, where $\rho$ and $\sigma$ are the number density and hard-sphere diameter, respectively. These results verify that Hadwiger's theorem from integral geometry, which predicts that $\gamma$ for a fluid at a surface, with certain restrictions, should be a linear combination of the average mean and Gaussian surface curvatures, is valid within the precision of the calculation for spherical and cylindrical surfaces up to $\eta \approx 0.42$. In addition, earlier results for $\gamma$ for this system [Bryk, {\em et al.}, Phys. Rev. E {\bf 68} 031602 (2003)] using a geometrically-based classical Density Functional Theory are in excellent agreement with the current simulation results  for packing fractions in the range where Hadwiger's theorem is valid. However, above $\eta \approx 0.42$, $\gamma(R)$ shows significant deviations from the Hadwiger form indicating limitations to its use for high-density hard-sphere fluids.   Using the results of this study together with Hadwiger's theorem allows one, in principle, to determine $\gamma$ for any sufficiently smooth surface immersed in a hard-sphere fluid. 
\end{abstract}
\maketitle

The solid-liquid interfacial free energy,  $\gamma$, is a central property governing a wide variety of  technologically important phenomena from crystal nucleation and growth to wetting. Because accurate and reliable experimental measurements of $\gamma$ are rare, much effort has been devoted in recent years to the development of atomistic simulation methods to determine this quantity for interfaces between coexisting solid and fluid phases\cite{Broughton86c,Davidchack00,Hoyt01,Laird05,Feng06} and for systems in which the solid is modeled by a static wall.\cite{Heni99,Miguel06,Fortini06,Fortini07,Laird07,Laird10}  These efforts have thus far been primarily restricted to planar interfaces; however, there are many physically relevant systems in which interfacial curvature is relevant, for example, in the formation of critical nuclei in nucleation\cite{Cacciuto03,Wu04,Bai06} or the  solvation/wetting of hydrophobic nanoscale particles.\cite{Lum99,Henderson02,Jin12} There have been a number of previous simulation studies that examine the effect of curvature in liquid-vapor interfaces\cite{Thompson84,Nijmeijer92,Block10,Troester12}, but direct simulation studies on solid-liquid interfaces are lacking.  In this work, we examine the dependence of $\gamma$ on the surface curvature for a hard-sphere fluid in contact with  curved hard surfaces, specifically at spherical and cylindrical colloidal particles. 

K\"{o}nig,  {\it et al.}\cite{Koenig04} have recently shown that Hadwiger's theorem\cite{Hadwiger57} from integral geometry puts severe restrictions on the shape (curvature) dependence of the interfacial free energy. In their analysis, the interfacial free energy of an object with a surface $S$ is given by
\begin{equation} 
\gamma(S) = \gamma_0 +h \bar{H} +\kappa \bar{K}
\label{eq:Hadwiger}
\end{equation}
where $h$ and $\kappa$ are constants depending upon the thermodynamic state, but independent of the specific surface $S$. Here $\bar{H}$ and $\bar{K}$ are the averaged mean and Gaussian curvatures of $S$, defined as
\begin{equation}
\bar{H} = \frac{1}{2 A} \int_{S} \left [ \frac{1}{R_1(\vec{q})} + \frac{1}{R_2(\vec{q})} \right ]  dS
\end{equation}
\begin{equation}
\bar{K} = \frac{1}{A} \int_{S} \left [ \frac{1}{R_1(\vec{q})\cdot R_2(\vec{q})}  \right ]  dS
\end{equation}
where $R_1(\vec{q})$ and $R_2(\vec{q})$ are the principal curvatures at each point $\vec{q}$ on the surface $S$, and $A$ is the surface area. As discussed in Ref.~\cite{Koenig04}, the use of Hadwiger's theorem to determine the free energy of fluids at bounding surfaces, often referred to as ``morphological thermodynamics", is valid as long as the fluid/surface system satisfies motion invariance, continuity and additivity. These conditions can break down for situations in which the bounding surface is small (i.e. on the order of the fluid particle size), for systems with long-range interactions, or for highly concave surfaces in which the fluid is confined in regions smaller than a few correlation lengths. For a fluid in contact with a spherical convex surface of radius $R$,  $\bar{H}_s = 1/R$ and $\bar{K}_s= 1/R^2$, while for a fluid in contact with at convex cylinder of the same radius, one has $\bar{H}_c = 1/2R$ and $\bar{K}_c = 0$. Much of the previous work on the curvature dependence focused on the first-order curvature correction,\cite{Tolman49} referred to as the Tolman length $\delta$ which, for a convex spherical surface, would be given in terms of $h$ by $\delta \gamma_0 = -h/2$. 

The first theoretical treatment for the curvature dependence of $\gamma$ for a hard-sphere fluid at a hard wall was developed within the so-called Scaled Particle Theory (SPT),\cite{Reiss60,Reiss65,Henderson02}  which is a theory of solvation that is based on an approximate determination of the work required to insert a spherical cavity (or hard-sphere solute) into a fluid.  For this system, $\gamma$ scales trivially with $T$, that is, $\gamma(\eta,T) = \gamma^*(\eta) kT/\sigma^2$, where $\gamma^*$ is the reduced interfacial free energy, $k$ is Boltzmann's constant, $T$ is temperature and $\sigma$ is the hard-sphere diameter of the fluid. The SPT result for  $\gamma^*$ for a hard-sphere fluid at a  convex spherical surface of radius $R$ is given by
\begin{equation}
\gamma^*_\mathrm{SPT}(\eta) = \frac{3\eta(2+\eta)}{2\pi(1-\eta)^2}+ \frac{3\eta}{2\pi(1-\eta)}\frac{1}{R}  +  \frac{\mathrm{ln}(1-\eta)}{4\pi}\frac{1}{R^2}
\label{eq:SPT-s}
\end{equation}
where $\eta$ is the packing fraction, defined in terms of the single particle density $\rho$ as $\eta = \pi \rho \sigma^3/6$. (For simplicity in what follows we measure distance in units of $\sigma$ and energy in units of $kT$ and drop the * superscript for reduced units.) The SPT treatment is consistent with the Hadwiger form of $\gamma$ (Eq.~\ref{eq:Hadwiger}) with $h$ and $\kappa$ equal to the coefficients of $1/R$ and $1/R^2$  in Eq.~\ref{eq:SPT-s}, respectively.  Note that, the value of the interfacial free energy between a fluid and a wall depends upon the precise definition of the dividing surface that determines the volume of the bounding wall.  In this work, we define the dividing surface to be coincident with the wall surface 

Using classical density functional theory (DFT), Bryk, {\it et al.}\cite{Bryk03} have examined  the scaling of the interfacial free energy and the excess interfacial adsorption  for  hard-sphere fluid at both spherical and cylindrical hard surfaces. This work, which utilized the Fundamental Measure Theory (FMT) version of DFT proposed by Rosenfeld\cite{Rosenfeld89}, determined $\gamma$ for this system as a function of packing fraction and curvature  with $1/R$ ranging from 0 to 0.5.  The results for the spherical and cylindrical geometries were fit to a polynomial of the form $\gamma(R) = \gamma_0 + a_1/R + a_2/R^2$ and were consistent with Hadwiger's theorem in that the ratio of $a_1$ for the spherical surface to that of the cylinder was 2. The predicted FMT values for $h$ for both geometries are in good agreement with the SPT results (Eq.~\ref{eq:SPT-s}) at low packing fractions, but exhibit a negative deviation from the SPT curve of a few percent at the higher densities studied. More recently, Jin, {\em et al.}\cite{Jin12} examined the solvation free energy of various ideal nanoscale non-spherical shapes (for example, cones, cylinders and prisms) using a hybrid Monte Carlo-DFT technique. They conclude that the use of "morphological thermodynamics" based on Hadwiger's theorem to determine  the solvation free energy (which includes both bulk and interfacial free energy components) for the non-spherical particles compares well with the DFT results.  To date, there have been no determinations of $\gamma$ from direct simulation for the hard-sphere fluid at curved surfaces with which to assess the accuracy of these DFT results. This is the principal goal of the present study. 

In this work, we make use of an adsorption equation derived using Cahn's extension\cite{Cahn79} of the surface thermodynamics of Gibbs\cite{Gibbs57}, namely, 
\begin{equation}
\left (\frac{\partial \gamma}{\partial P}\right )_T = v_N
\label{eq:gibscahn}
\end{equation} 
where the excess interfacial volume per unit area, $v_N$, is defined by
\begin{equation}
v_N =   \frac{1}{A\,N_f} \left | \begin{array}{cc} V&N\\V_f & N_f \end{array} \right | =  \frac{1}{A}\left (V - V_f\frac{N}{N_f} \right )
\label{eq:vndef}
\end{equation}
where $A$ is the interfacial area,  $V$ and $N$ are the volume and number of particles, respectively,  of a region containing the interface and $V_f$ and $N_f$ are the corresponding quantities for a region entirely within the bulk fluid.\cite{Laird10} This adsorption equation can be shown to be equivalent through a change of variables and Maxwell relations to the usual Gibbs adsorption equation
\[
\left (\frac{\partial \gamma}{\partial \mu}\right )_T = \Gamma_N 
\]
where $\Gamma_N$ is the excess interfacial number of particles per unit area at the interface. Eq.~\ref{eq:gibscahn} can be integrated with respect to pressure to give
\begin{equation}
\gamma(P) = \gamma_{P=0} + \int_0^{P} v_N(P)\, dP
\label{eq:gamma_hs_P}
\end{equation}
where $P$ is the pressure. For the hard-sphere/hard-wall case, $\gamma_{P = 0} = 0$. Eq.~\ref{eq:gamma_hs_P} was recently used to determine $\gamma$ for a hard-sphere fluid at a flat wall\cite{Laird10}. The excess interfacial volume can be related to the density profile $\rho(r)$ by
\begin{equation}
v_N = \frac{1}{R^i} \int_R^{\infty} \left [ 1 - \frac{\rho(r)}{\rho_f} \right ] r^i \,dr
\label{eq:vn_rho}
\end{equation}
where $i = 1$ and 2 for the cylindrical and spherical geometries, respectively. 

To determine $\gamma$ for this system using Eq.~\ref{eq:gamma_hs_P}, we use molecular-dynamics (MD) simulation to calculate $v_N$ as a function of bulk packing fraction $\eta = \rho\sigma^3/6$ for a hard-sphere fluid in contact with spherical and cylindrical colloidal particles of radius $R$  varying from 10 down to 0.5, over a range of  $\eta$ from 0.03 to 0.49, the upper limit being the fluid packing fraction at freezing. For the simulations, we use the algorithm of Rapaport \cite{Rapaport95} and use Eq.~\ref{eq:vndef} directly to determine $v_N$ in the simulations. For additional simulation details and plots of $v(\eta;R)$ see the Supplemental Information.\cite{supplemental}  

Because we find $v_N$ as a function of the packing fraction, not the pressure, we transform Eq.~\ref{eq:gamma_hs_P} to give
\begin{equation}
\gamma(\eta) = \int_0^{\eta} v_N(\eta^{\prime}) \left ( \frac{\partial P}{\partial \eta^{\prime}}\right )_T\, d\eta^{\prime}
\label{eq:gamma_hs_eta}
\end{equation}
To obtain the derivative of $P$ with respect to $\eta$ we use the KLM-low equation of state (EOS).\cite{Kolafa04} This EOS has been shown to give five decimal place accuracy in the pressure even at high density when compared to high-quality simulations,\cite{Bannerman10} so any errors introduced by its use are much smaller than the statistical error in the simulation data. 
To reduce the numerical integration error in evaluating Eq.~\ref{eq:gamma_hs_eta}  we subtract from the integrand the corresponding value obtained from the SPT (Eq.~\ref{eq:SPT-s}) - the corresponding SPT expression for the cylindrical surface is obtained from 
 Eq.~\ref{eq:SPT-s} by setting the coefficient of the $1/R^{2}$ term to zero and dividing the  coefficient of the $1/R$ term by two. Accurate calculation of the excess  interfacial volume $v$ at very low densities is difficult because of sampling issues; however, this quantity can be calculated exactly in the limit $\rho \rightarrow 0$. In this limit, we have 
 \begin{equation}
\lim_{\rho \rightarrow 0} v_N(R) = \frac{1}{2} + \frac{a_1}{R} + \frac{a_2}{R^{2}}  \end{equation}
 where $a_1$ and $a_2$ are constants equal to 1/4 (1/8) and 1/24 (0), respectively, for the spherical (cylindrical) wall.

For small values of $R$, the statistical errors in $v_N$ are larger for similar simulation lengths because the number of particles near the wall is relatively small.  For the special case of $R = 0.5$ where the spherical surface is identical in size to the fluid particles, high precision results can be generated by recognizing that the density profile $\rho(r)$ is equivalent to  $\rho_f g(r)$, where $g(r)$ is the radial distribution in the bulk hard sphere fluid.  For this value of  $R$, we use an MD simulation for a bulk hard-sphere fluid to calculate $g(r)$ from which $v_N$ can be determined with high precision. Alternatively, by replacing $\rho(r)$ with $\rho_f g(r)$ for $R = 0.5$, we have from Eq.~\ref{eq:vn_rho} that
\begin{eqnarray} 
v_N (R = 0.5) & =& 4 \int_{1/2}^{\infty} r^2 [1-g(r)] dr \\ \nonumber
 &= &-\frac{1}{\pi} \int_{0}^{\infty}4 \pi r^2  [g(r) - 1]  dr - 1/6
\end{eqnarray} 
Using the compressibility equation from liquid state physics,\cite{Hansen06} the integral in the previous equation can be replaced with a  term dependent upon the isothermal compressibility $\kappa_T = \rho^{-1} (\partial \rho/\partial P)_T$ to yield
\begin{equation}
v_N(\rho; R_s = 1/2) = -\frac{1}{\pi} (kT \kappa_T - \rho^{-1}) - 1/6
\end{equation}
Using an equation of state, the isothermal compressibility $\kappa_T$ can be determined analytically as a function of $\eta$ allowing for an analytical calculation of $v_N$, and thus $\gamma$ for $R = 1/2$.

\begin{figure}[h]
\includegraphics[width= \columnwidth]{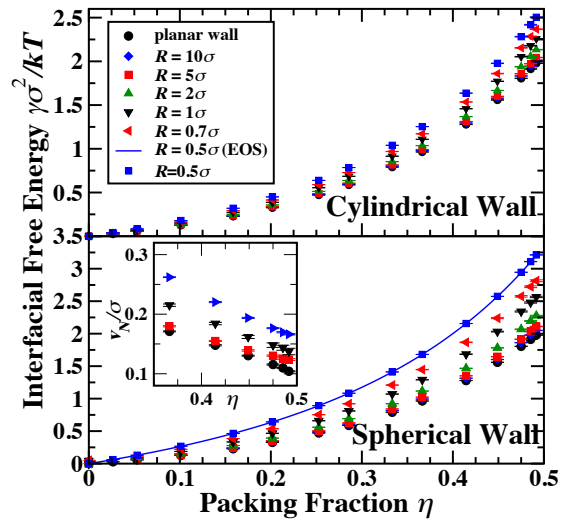}
\caption{Color online. Top panel: Interfacial free energy between a hard-sphere fluid and a hard cylindrical colloidal particle as a function of $\eta$ for several values of the radius, $R$. Bottom panel: Same as the top panel except for the spherical surface. The inset  shows the value of the excess volume, $v_N$, for the spherical surface as a function of packing fraction for $R = \infty, 5$ and 0.5. The symbols in this panel and in the inset are as indicated in the top panel legend.}
  \label{fig:gamma}
\end{figure}

The calculated values of $\gamma$ from our simulations for the spherical and cylindrical walls are plotted as functions of $\eta$ in Fig.~\ref{fig:gamma}, for values of $R$  ranging from the planar wall ($R = \infty$) to 0.5. At all densities, these figures show that at fixed $\eta$, the interfacial free energy $\gamma$ is a monotonically increasing function of $1/R$. 

\begin{figure}[h]
 \includegraphics[width= \columnwidth]{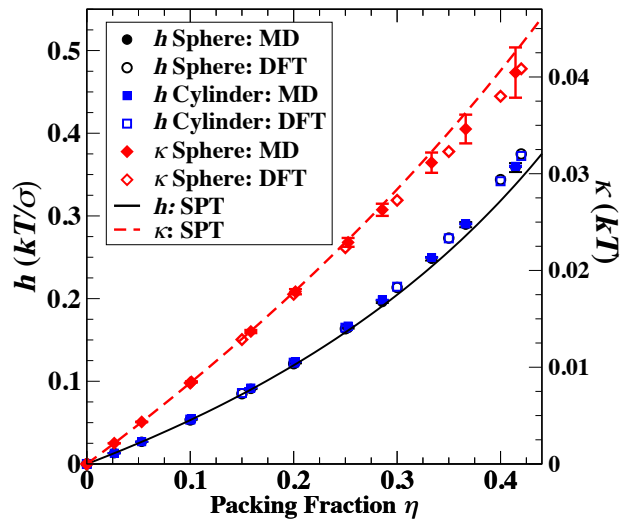}
\caption{Color online. The calculated values of $h$ and $\kappa$ for the spherical and cylindrical walls. Filled symbols are the results from the present MD simulations and the open symbols are the values calculated from the White-Bear DFT. The solid and dashed lines are the values of $h$ and $\kappa$ from Scaled Particle Theory (SPT). Note that, the symbols for the various values of $h$ are difficult to resolve from one another because they are nearly coincident. }
  \label{fig:h_kappa}
\end{figure}

To test the validity of Hadwiger's theorem for this system, we fit the data shown in Fig.~\ref{fig:gamma}  to Eq.~\ref{eq:Hadwiger} to determine the constants $h$ and $\kappa$ for each $\eta$. The fit was 
performed using standard weighted quadratic (for the spherical case where $\bar{H} = 1/R$ and $\bar{K} = 1/R^2$) and 
linear (for the cylindrical case where $\bar{H} = 1/2R$ and $\bar{K} = 0$ ) least-squares regression. The weights in the 
regression were equal to the inverse of the statistical variance of the data points. The results of these fits are shown in 
Fig.~\ref{fig:h_kappa} for packing fractions up to 0.42.
A table of the fitted values for $\gamma_0$, $h$ and $\kappa$ is included in the Supplemental Information.\cite{supplemental} Also, shown in 
Fig.~\ref{fig:h_kappa} are the results for $h$ and $\kappa$ from  the SPT expression~\ref{eq:SPT-s} and from the 
Rosenfeld DFT.\cite{Bryk03}  The DFT results  agree remarkably  well with the simulation results for the values of $\eta$ shown in Fig.~\ref{fig:h_kappa}. Most notable in these  results is the fact that the estimates for $h$ for the spherical and cylindrical geometries shown in Fig.~\ref{fig:h_kappa} are in excellent agreement, as predicted by Hadwiger's theorem. The SPT prediction agrees well with the simulation results at low packing fractions, but  underestimates $h$ and slightly overestimates $\kappa$ at packing fractions above about 0.25.  Note that, as mentioned earlier, the value of $\gamma$ and, by extension, the Hadwiger coefficients $\gamma_0$, $h$ and $\kappa$ will depend upon the choice of dividing surface; however, the Hadwiger form (Eq.~\ref{eq:Hadwiger}) is remains valid, albeit with modified coefficients.\cite{Henderson02}

\begin{figure}[h]
  \includegraphics[width= \columnwidth]{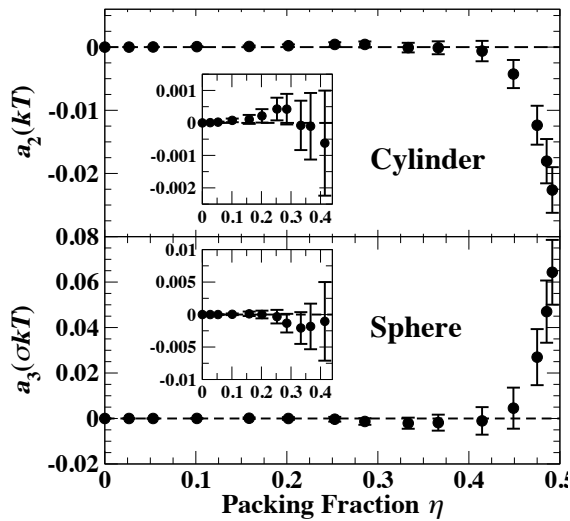}
\caption{Bottom panel:  Plot of the first non-Hadwiger coefficient (cubic, $a_3$) of $1/R$  in the curvature expansion of $\gamma$  for the hard-sphere fluid at a spherical colloidal particle as a function of packing fraction $\eta$. For clarity, the inset shows the data for $\eta < 0.42$ on a smaller scale. Top panel: Same as bottom panel, but for the cylindrical surface.}
  \label{fig:nonHad}
\end{figure}

Above $\eta = 0.42$, however, significant deviations from the Hadwiger form were found. To quantify this deviation, we performed a 
cubic and quadratic weighted least-square regression with respect to $1/R$ on the sphere and cylinder data, respectively. The results of these calculation are plotted in Fig.~\ref{fig:nonHad}. At packing fractions below 0.42, the cubic coefficient ($a_3$) for the spherical  surface  is zero within the estimated statistical error, consistent with Hadwiger's theorem, but this coefficient diverges quickly from zero at packing fractions above 0.42, indicating a significant breakdown of the Hadwiger form at packing fractions approaching the freezing density ($\eta = 0.492$).  The situation is similar for the quadratic coefficient ($a_2$) for the cylindrical case (Fig.~\ref{fig:nonHad} top panel)- there is one value of $a_2$ at $\eta ~ 0.25$ that is marginally different from zero outside the error bars, but given its marginality and the small value of the coefficient, we do not view this deviation as significant.  The origin of the divergence of the non-Hadwiger  coefficients at high packing fraction can be seen in the inset in Fig.~\ref{fig:gamma} in which the excess volume, $v_N$, is plotted for the spherical surface as a function of $\eta$ for $R/\sigma = \infty, 5, 1$ and 0.5. At high $\eta$, the excess volume $v_n$  for the planar wall ($R = \infty$) exhibits a downward curvature that is not present in the other values of $R$, making it impossible to fit $v_N(R)$ purely as a quadratic polynomial in $1/R$.  It is possible that this  anomalous decrease in $v_N$  for the planar wall is connected to the prefreezing transition that has been observed for the hard-sphere fluid at a 
hard planar wall at high packing fraction\cite{Courtemanche92, Dijkstra04,Laird07}, and further analysis is underway. 
These results also indicate that the errors in the SPT at the intermediate packing fractions are not likely to be due to the omission of higher-order terms (beyond the quadratic for the sphere) in $1/R$, as has been proposed (see Ref.~\onlinecite{Ashbaugh06} for a review). The results here put severe upper bounds on the magnitude of any such higher-order terms. As an example, our calculated value for $\gamma$ at $\eta = 0.25265$ and $R = 1.0$ is 0.6595(6). The corresponding SPT value from Eq.~\ref{eq:SPT-s} is 0.6711 -  a difference of 0.0116(6), which is more than an order of magnitude larger than the estimated upper bound to the contribution of the cubic term $a_3$ at this packing fraction from Fig.~\ref{fig:nonHad}. Note that largest contribution to the difference comes from $\gamma_0$, the value for the planar wall. 

In summary, we have calculated the curvature dependence of the interfacial free energy between hard-sphere fluid and 
hard spherical and cylindrical colloidal particles of varying radius, $R$. These calculations have important applications in determining the solvation free energy of nanoscale objects.  From our simulation results, we have verified 
that the predictions of Hadwiger's theorem,\cite{Hadwiger57,Koenig04} which predicts that $\gamma$ for fluids at a  bounding surface, under certain restrictions, is completely described by a linear combination of the average mean and  Gaussian curvatures,  are valid for the systems studied at low to moderate fluid packing fractions 
($\eta < 0.42$) and have calculated the coefficients of this linear combination ($h$ and $\kappa$) as functions  $\eta$. 
 For these packing fractions, the Hadwiger form for $\gamma$ (Eq.~\ref{eq:Hadwiger}) is shown to be valid for  both convex spherical and cylindrical surfaces even down to $R = 0.5\sigma$, where the radius of the bounding surface is  equal to that of the fluid particles. In addition, earlier DFT results based on Rosenfeld's fundamental measure theory are 
found to be in excellent agreement with the results of the current simulations.  At higher packing fractions $(\eta > 0.42)$, however, significant deviations from the Hadwiger form are observed. The  results of these simulations can serve as a useful reference model to determine the thermodynamics of the hydrophobic solvation of nanoscale particles.

\section{Acknowledgments} \label{sec:acknw}
BBL acknowledges support from the National Science Foundation (NSF) under grant CHE-0957102. We also acknowledge the KU Research Experiences for Undergraduates funded by NSF under grant CHE-1004897 for providing AH with support to work on this project.  This research used the ALICE High Performance Computing Facility at the University of Leicester. We also wish to thank Dr. Roland Roth for helpful discussions.


\begin{thebibliography}{10}

\bibitem{Broughton86c}
J.Q. Broughton and G.H. Gilmer, J. Chem. Phys. {\bf 84}, 5759--5768 (1986).

\bibitem{Davidchack00}
R.L. Davidchack and B.B. Laird, Phys. Rev. Lett. {\bf 85}, 4751--4754 (2000).

\bibitem{Hoyt01}
J.J. Hoyt, M.~Asta, and A.~Karma, Phys. Rev. Lett {\bf 86}, 5530--5533 (2001).

\bibitem{Laird05}
B.B. Laird and R.L. Davidchack, J. Phys. Chem. B {\bf 109}, 17802
  (2005).

\bibitem{Feng06}
X.~Feng and B.B. Laird, J. Chem. Phys. {\bf 124}, 044707 (2006).

\bibitem{Heni99}
M.~Heni and H.~L{\"o}wen, Phys. Rev. E {\bf 60}, 7057 (1999).

\bibitem{Miguel06}
E.~De Miguel and G.~Jackson, Mol. Phys. {\bf 104}, 3717 (2006).

\bibitem{Fortini06}
A.~Fortini and M.~Dijkstra, J. Phys.: Condens. Matter {\bf 18}, L371
  (2006).

\bibitem{Fortini07}
A.~Fortini.
\newblock PhD thesis, Utrecht University, The Netherlands (2007).

\bibitem{Laird07}
B.B. Laird and R.L. Davidchack, J. Phys. Chem. C {\bf 111}, 15952
  (2007).

\bibitem{Laird10}
B.B. Laird and R.L. Davidchack, J. Chem. Phys. {\bf 132}, 204101 
  (2010).
  
\bibitem{Cacciuto03}
A. Cacciuto, S. Auer and D. Frenkel,  J. Chem. Phys. {\bf 119},  7467 (2003).

\bibitem{Wu04}
D.T. Wu, L. Granasy and F. Spaepen, MRS Bullet. {\bf 29}, 945 (2004). 

\bibitem{Bai06}
X.-M. Bai and M. Li, J. Chem. Phys. {\bf 124}, 124707 (2006).

\bibitem{Lum99}
K. Lum, D. Chandler and J.D. Weeks, J. Phys. Chem. {\bf 103}, 4570 (1999). 

\bibitem{Henderson02}
J.~R. Henderson, J. Chem. Phys. {\bf 116}, 5039 (2002).

\bibitem{Jin12}
Z. Jin, J. Kim and J. Wu, Langmuir, {\bf 28}, 6997 (2012). 

\bibitem{Thompson84}
S.M. Thompson, K.E. Gubbins, J.P.R.B. Walton, R.A.R. Chantry and J.S. Rowlinson, J. Chem. Phys. {\bf 81} 530 (1984).

\bibitem{Nijmeijer92}
M.J.P. Nijmeijer, C. Bruin, A.B. van Woerkom, A.F. Bakker and J.M.J. van Leeuwen, J.Chem. Phys. {\bf 96} 565 (1992). 

\bibitem{Block10}
B.J. Block, S.J. Das, M. Oettel, P. Virnau and K. Binder, J. Chem. Phys. {\bf 133}, 154702 (2010). 

\bibitem{Troester12}
A. Tr\"{o}ster, M. Oettel, B. Block, P. Virnau and K. Binder, J. Chem. Phys. {\bf 136}, 064709 (2012). 

\bibitem{Koenig04}
P.M. K\"{o}nig, R.~Roth, and K.R. Mecke, Phys. Rev. Lett. {\bf 93}, 160601
  (2004).

\bibitem{Hadwiger57}
H.~Hadwiger, {\em Vorlesung {\"u}ber Inhalt Oberfl{\"a}che und Isoperimetrie},
  (Springer, Berlin, 1957).

\bibitem{Tolman49}
R.C. Tolman, J. Chem. Phys. {\bf 17}, 333 (1949).

\bibitem{Reiss60}
H.~Reiss, H.L. Frisch, E.~Helfand, and J.L. Lebowitz, J. Chem. Phys. {\bf 32},
  119 (1960).

\bibitem{Reiss65}
H.~Reiss, Adv. Chem. Phys. {\bf 9}, 1 (1965).

\bibitem{Bryk03}
P.~Bryk, R.~Roth, K.R. Mecke, and S.~Dietrich, Phys. Rev. E {\bf 68}, 031602
  (2003).

\bibitem{Rosenfeld89}
Y.~Rosenfeld, Phys. Rev. Lett. {\bf 63}, 980 (1989).

\bibitem{Cahn79}
J.W. Cahn in {\em Interfacial
  Segregation}, W.C. Johnson and J.M. Blakely, eds., (ASM International, International Materials Park,
  OH, 1979), pp. 3--23

\bibitem{Gibbs57}
J.W. Gibbs, {\em The Collected Works}, Vol.~1, (Yale University Press, New
  Haven, 1957).

\bibitem{Rapaport95}
D.~C. Rapaport, {\em The Art of Molecular Dynamics Simulation}, (Cambridge
  University Press, New York, 1995).
  
\bibitem{supplemental} 
See Supplemental Material at [URL will be inserted by publisher] for additional simulation details and intermediate results. 
\bibitem{Kolafa04}
J.~Kolafa, S.~Lab\'{i}k, and A.~Malijevsk\'{y}, Phys. Chem. Chem. Phys. {\bf
  6}, 2335 (2004).

\bibitem{Bannerman10}
M.N. Bannerman, L.~Lue, and L.V. Woodcock, J. Chem. Phys. {\bf 132}, 084507 (2010).

\bibitem{Hansen06}
J.P. Hansen and I.R. McDonald, {\em Theory of Simple Liquids, 3rd Ed.}, (Academic Press,
  New York, 2006).

\bibitem{Courtemanche92}
D.J.  Courtemanche,  F. van Swol, Phys. Rev. Lett.   {\bf 69}, 2078 (1992). 

 \bibitem{Dijkstra04} M. Dijkstra,  Phys. Rev. Lett.  {\bf 93}, 108303 (2004). 

\bibitem{Ashbaugh06} 
H.S. Ashbaugh and L.R. Pratt, Rev. Mod. Phys. {\bf 78}, 159 (2006).

\end{thebibliography}
\end{document}